\documentclass[10pt]{article}
\usepackage{fancyhdr}
\usepackage{extramarks}
\usepackage{amsmath}
\usepackage{amsthm}
\usepackage{amsfonts}
\usepackage{siunitx}
\usepackage{tikz}
\usepackage[plain]{algorithm}
\usepackage{algpseudocode}
\usepackage{multirow}
\usepackage{booktabs}
\usepackage{graphicx}
\usepackage{subfigure}
\usepackage[colorlinks,linkcolor=black,anchorcolor=black,citecolor=black,urlcolor=blue]{hyperref}
\usepackage{amsmath,bm}
\usepackage{booktabs}
\usepackage{mathtools}
\usepackage{amssymb}
\usepackage{caption}
\usepackage{capt-of}
\usepackage{mciteplus}
\usepackage{cite}
\usepackage{mathrsfs}
\usepackage[title,titletoc,toc]{appendix}
\usepackage{xr}
\usepackage{parskip}
\usepackage{soul}
\usepackage{textcomp}
\usepackage[colaction]{multicol}
\usepackage[switch]{lineno}
\usepackage{lipsum}
\usepackage{etoolbox}
\usepackage{longtable}
\usepackage{array}
\usepackage{tablefootnote}
\usepackage{ragged2e}
\newcolumntype{C}[1]{>{\centering\arraybackslash}p{#1}}
\usetikzlibrary{automata,positioning}
\usetikzlibrary{automata,positioning}

\begin{document}

\title{Omicron (B.1.1.529): Infectivity,  vaccine breakthrough, and antibody resistance.    
} 
\author{ Jiahui Chen$^1$, Rui Wang$^1$, Nancy Benovich Gilby$^2$ and Guo-Wei Wei$^{1,3,4}$\footnote{
 		Corresponding author.		Email: weig@msu.edu} \\
 $^1$ Department of Mathematics, \\
 Michigan State University, MI 48824, USA.\\
 $^1$ Spartan Innovations, \\
325 East Grand River Ave., Suite 355,\\
East Lansing, MI 48823 USA.\\
 $^3$ Department of Electrical and Computer Engineering,\\
 Michigan State University, MI 48824, USA. \\
 $^4$ Department of Biochemistry and Molecular Biology,\\
 Michigan State University, MI 48824, USA. \\
 }
\date{\today} 

\maketitle

\begin{abstract}
The latest severe acute respiratory syndrome coronavirus 2 (SARS-CoV-2) variant Omicron (B.1.1.529) has ushered panic responses around the world due to its contagious and vaccine escape mutations. The essential infectivity and antibody resistance of the SARS-CoV-2 variant are determined by its mutations on the spike (S) protein receptor-binding domain (RBD). However, a complete experimental evaluation of Omicron might take weeks or even months. Here, we present a comprehensive quantitative analysis of Omicron's infectivity, vaccine-breakthrough, and antibody resistance. An artificial intelligence (AI) model, which has been trained with tens of thousands of experimental data points and extensively validated by experimental data on SARS-CoV-2, reveals that Omicron may be over ten times more contagious than the original virus or about twice as infectious as the Delta variant. Based on 132 three-dimensional (3D) structures of antibody-RBD  complexes, we unveil that   Omicron may be twice more likely to escape current vaccines than the Delta variant. The Food and Drug Administration (FDA)-approved monoclonal antibodies (mAbs) from Eli Lilly may be seriously compromised. Omicron may also diminish the efficacy of mAbs from Celltrion and Rockefeller University. However, its impact to Regeneron mAb cocktail appears to be mild.

\end{abstract}
\newpage

\setcounter{page}{1}
\renewcommand{\thepage}{{\arabic{page}}}


%
 
\section{Introduction}
On November 26, 2021, the World Health Organization (WHO) announced a new severe acute respiratory syndrome coronavirus 2 (SARS-CoV-2) variant Omicron (B.1.1.529), as a variant of concern (VOC). This variant carries an unusually high number of mutations, 32, on the spike (S) protein, the main antigenic target of antibodies generated by either infections or vaccination. In comparison, the devastating Delta variant has only 5 S protein mutations, which posed a high potential global risk and has spread internationally. Therefore, the ``panic button" has been pushed in several cases worldwide, and many countries have enacted travel restrictions to prevent the rapid spread of the Omicron variant. 

The mutations on the Omicron variant are widely distributed on multiple proteins of SARS-CoV-2 such as NSP3, NSP4, NSP5, NSP6, NSP12, NSP14,  S protein, envelope protein, membrane protein, and nucleocapsid protein. The focus is the mutations on the S protein receptor-binding domain (RBD) for the potential impact on infectivity and antibody resistance caused by this new variant. This is due to the fact that the RBD located on the S protein facilitates the binding between the S protein and the host angiotensin-converting enzyme 2 (ACE2). Such S-ACE2 binding helps the SARS-CoV-2 enter the host cell and initiates the viral infection process. Several studies have shown that the binding free energy (BFE) between the S RBD and the ACE2 is proportional to the viral infectivity \cite{li2005bats,qu2005identification,song2005cross,hoffmann2020sars,walls2020structure}. As such, an antibody that binds strongly to the RBD would directly neutralize the virus \cite{wang2020human,yu2020receptor,li2021impact}. Indeed, many RBD binding antibodies are generated by the human immune response to infection or vaccination. Monoclonal antibodies (mABs) targeting the S protein, particularly the RBD, are designed to treat viral infection. As a result, any mutation on the S protein RBD would cause immediate concerns about the efficacy of existing vaccines, mAbs and the potential of reinfection.

The global panic brought by the emergence of the Omicron variant drives the scientific community to immediately investigate how much this new variant could undermine the existing vaccines and mAbs. However, relatively reliable experimental results from experimental labs will take a few weeks to come out. Therefore, an efficient and reliable in-silico analysis is imperative and valuable for such an urgent situation. Here, we present a comprehensive topology-based artificial intelligence (AI) model called TopNetmAb \cite{chen2020mutations, chen2021prediction} to predict the BFE changes of S and ACE2/antibody complexes induced by mutations on the S RBD of the Omicron variant. The positive BFE change induced by a specific RBD mutation indicates its potential ability to strengthen the binding of an S protein-ACE2/antibody complex, while a negative BFE change suggests a likely capacity to reduce the binding strength of an S protein-ACE2/antibody complex.

The TopNetmAb model that we proposed has been extensively validated over the past 1.9 years \cite{chen2021prediction,chen2021revealing}. Initially, in early 2020, we applied our TopNetmAb model to successfully predict that residues 452 and 501 "have high chances to mutate into significantly more infectious COVID-19 strains" \cite{chen2020mutations}. Such findings have been confirmed due to the emergency of multiple variants such as Alpha, Beta, Gamma, Delta, Theta, Lambda, Mu, and Omicron that carry L452R/Q and N501Y mutations. In April 2021, we provided a list of 31 RBD mutations that may weaken most of the binding to antibodies, such as W353R, I401N, Y449D, Y449S, P491R, P491L, Q493P \cite{chen2021prediction}. Notably, experimental results have also shown that mutations at residues Y449, E484, Q493, S494, Y505 might enable the virus to escape antibodies \cite{alenquer2021signatures}. Meanwhile, in the same work, we also revealed that variants found in the United Kingdom and South Africa in late 2020 may strengthen the binding of the RBD-ACE2 complex, which is consistent with the experimental results \cite{dupont2021neutralizing}. Later on, we provided a list of most likely vaccine escape RBD mutations predicted by TopNetmAb, such as S494P, Q493L, K417N, F490S, F486L, R403K, E484K, L452R, K417T, F490L, E484Q, and A475S \cite{wang2021vaccine}, and mutations such as S494P, K417N, E484K/Q, L452R are all detected in the variants of concern and variants of interest denounced by WHO. Last but not least, the correlation between the experimental deep mutational data \cite{linsky2020novo} and our AI-predicted RBD-mutation-induced BFE changes for all possible 3686 RBD mutations on the RBD-ACE2 complex is 0.7, which indicates the reliability of our TopNetmAb model predictions \cite{chen2021revealing}.

This work aims to analyze how the RBD mutations on the Omicron variant will affect the viral infectivity and efficacy of existing vaccines and antibody drugs. Fifteen Omicron RBD mutations, including S371L, S373P, S375F, K417N, N440K, G446S, S477N, T478K, E484A, Q493R, G496S, N501Y, and Y505H, are studied in this work. Additionally, three-dimensional (3D) structures of RBD-ACE2 complex and 132  antibody-RBD complexes, including many mAbs, are examined to understand the impacts of Omicron RBD mutations.  
We reveal that Omicron may be over ten times more contagious than the original SARS-CoV-2 and about twice as infectious as the Delta variant, mainly due to its RBD mutations N440K, T478K, and N501Y. Additionally, Omicron has a high potential to disrupt the binding of most 132 antibodies with the S protein, mainly due to its RBD mutations K417N, E484A, and Y505H,  indicating its stronger vaccine-breakthrough capability than the Delta variant. We have also unveiled that Omicron may seriously reduce the efficacy of the Eli Lilly mAb cocktail because of Omicron RBD mutations K417N, E484A, and Q493R. Celltrion antibody Regdanvimab may be disrupted by Omicron RBD mutations E484A, Q493R, and Q498R. Omicron RBD mutation E484A may also disrupt Rockefeller University mAbs. However, Omicron's  impact on the Regeneron mAb cocktail is predicted to be mild. 

\section{Results}

\subsection{Infectivity}

The Infectivity of SARS-CoV-2 is mainly determined by the binding affinity of the ACE2 and RBD complex, although the furin cleavage site plays a crucial role as well \cite{zhang2021furin}. 
Omicron has three mutations at the furin cleavage site and 15 mutations on the RBD, suggesting a significant change in its infectivity. Due to natural selection, the virus enhances its evolutionary advantages at the RBD either by mutations to strengthen the ACE2-RBD binding affinity or by mutations to escape antibody protection \cite{wang2021emerging}. Since the virus has optimized its Infectivity in human cells, one should not expect a dramatic increase in the viral infectivity by any single mutation. An effective infection pathway is for the virus to have multiple RBD mutations to accumulatively enhance its infectivity, which appears to be the case for Omicron.

\begin{figure}[ht!]
	\centering
	\includegraphics[width = 0.95\textwidth]{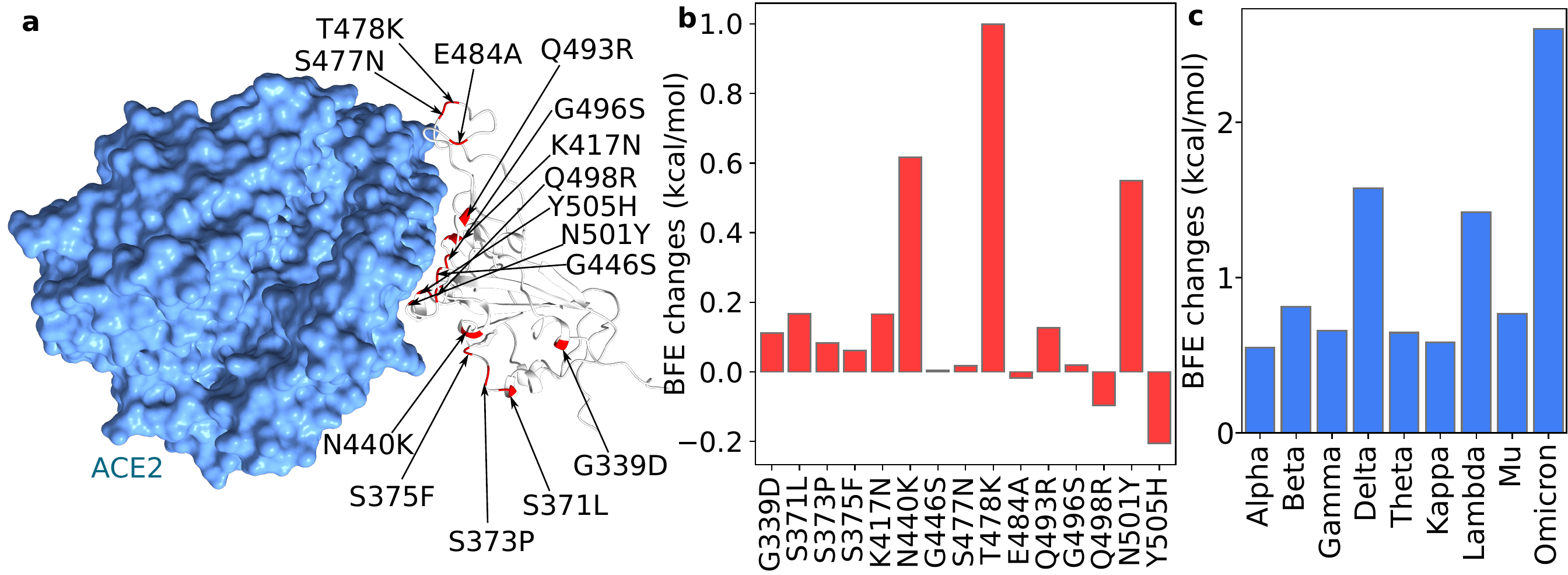}
	\caption{Illustration of the Omicron RBD and ACE2 interaction, RBD mutation-induced BFE changes.  
{\bf a.} The 3D structure of the ACE2 and RBD complex. Omicron mutation sites are labeled.  
{\bf b.} Omicron mutation-induced BFE changes.  Positive changes strengthen the binding between ACE2 and S protein, while negative changes weaken the binding. 
{\bf c.} A comparison of predicted mutation-induced BFE changes for few variants.
}
	\label{fig:combine1}
\end{figure}

This work analyzes the infectivity of Omicron by examining the BFE changes of the ACE2 and S protein complex induced by 15 Omicron RBD mutations. 
Figure \ref{fig:combine1}{\bf a} illustrates the binding complex of ACE2 and S protein RBD. Most of the RBD mutations are located near the binding interface of ACE2 and RBD, except for mutations G339D, S371L, S373P, and S375F.  Omicron-induced BFE changes are depicted in Figure \ref{fig:combine1}{\bf b}. Overall, mutations significantly increase the BFE changes, which strengthens the binding affinity of the ACE2-RBD complex and makes the variant more infectious. This result indicates that Omicron appears to have followed the infectivity-strengthening pathway of natural selection \cite{chen2021review2}.

The infectivity-strengthening mutations N440K, T478K, and N501Y enhance the BFEs by 0.62, 1.00,  and 0.55 kcal/mol, respectively. Among them, T478K is one of two RBD mutations in the Delta variant, while N501Y is presented on many prevailing variants, including Alpha, Beta, Gamma, Theta, and Mu. Notably, mutation Y505H induces a small negative BFE change of -0.20kcal/mol. All other mutations, particular those four mutations that are far away from the ACE2 and RBD binding interface,  cause little or no BFE changes. Figure \ref{fig:combine1}{\bf c}  gives a comparison of Omicron with a few other variant.
Overall, the accumulated BFE change is 2.60kcal/mol, suggesting a 13-fold increase in the viral Infectivity. In comparison, Omicron is about 2.8 times as infectious as the Delta (i.e., BFE change: 1.57kcal/mol).

\subsection{Vaccine breakthrough}

Vaccination has been proven to be the most effective means for COVID-19 prevention and control. There are four types of vaccines, i.e., virus vaccines, viral-vector vaccines, DNA/RNA vaccines, and protein-based vaccines \cite{callaway2020race}. Essentially,  the current COVID-19 vaccines in use mainly target to the S protein \cite{dai2021viral}. The 32 amino acid changes, including three small deletions and one small insertion in the spike protein, suggest that Omicron may  be induced by vaccination. As a result, these mutations may dramatically enhance the variant's ability to evade current vaccines. 

In general, it is essentially impossible to accurately characterize the full impact of Omicron's S protein mutations on the current vaccines in the world's populations. First, different types of vaccines may lead to different immune responses from the same individual. Additionally, different individuals characterized by race, gender, age, and underlying medical conditions may produce different sets of antibodies from the same vaccine. Moreover, the reliability of statistical analysis over populations may be limited because of the inability to fully control various experimental conditions. 

This work offers a molecule-based data-driven analysis of Omicron's impact on vaccines through a library of 132 known antibody and S protein complexes. 
We evaluate the binding free energy changes induced by 15 RBD mutations on these complexes to understand the potential impact of Omicron's RBD mutations to vaccines. To ensure reliability, our study does not include a few known antibody-S protein complexes that are far away from the RBD, such as those in the N-terminal domain (NTD), due to limited experimental data in our antibody library \cite{chen2021prediction,chen2021revealing}.

\begin{figure}[ht!]
	\centering
	\includegraphics[width = 1\textwidth]{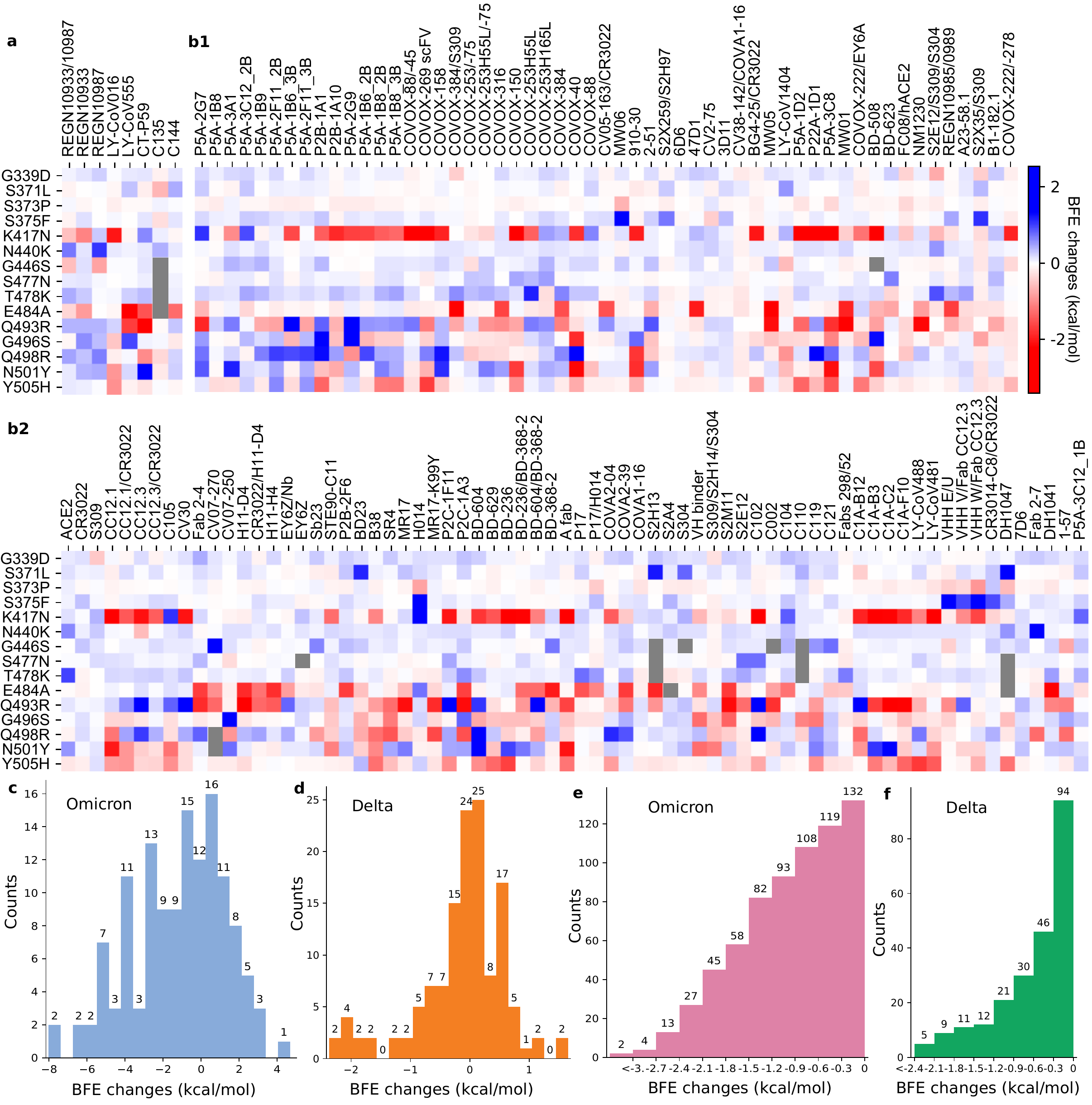}
\caption{Illustration of Omicron mutation-induced BFE changes of 132 available antibody and RBD complexes. Positive changes strengthen the binding, while negative changes weaken the binding. 
{\bf a} Heat map for 8 antibodies and RBD complexes in various stages of drug development. Gray color stands for no predictions due to incomplete structures. 
{\bf b1-b2} Heat map for 124 antibodies and RBD complexes. 
{\bf c} The distribution (counts) of accumulated BFE changes  induced by Omicron mutations for 132 antibody and RBD complexes. Overall, there are more complexes that are weakened upon Omicron RBD mutations that complexes that are strengthened.
 {\bf d} The distribution (counts) of accumulated BFE changes  induced by Delta mutations for 132 antibody and RBD complexes. 
 {\bf e} The numbers (counts) of antibody and RBD complexes regarded as disrupted by Omicron under different thresholds ranging from 0 kcal/mol, -0.3 kcal/mol, to $<$-3 kcal/mol. 
{\bf f} The numbers (counts) of antibody and RBD complexes regarded as disrupted by Delta under different thresholds ranging from 0 kcal/mol, -0.3 kcal/mol, to $<$-2.4 kcal/mol. 
}
	\label{fig:combine2}
\end{figure}

Figures \ref{fig:combine2}{\bf a, b1}, and {\bf b2} depict the Omicron RBD mutation-induced  BFE changes of 132 known antibody and RBD complexes.
Overall, Omicron RBD mutations can significantly change the binding pattern of known antibodies. Positive changes strengthen the binding between antibody and RBD complexes, while negative changes weaken the binding. In the color bar,  the largest negative change is more significant than the largest positive change, indicating more severe disruptive impacts.  In general, there are more negative BFE changes than positive ones, as shown in Figure  \ref{fig:combine2}{\bf c}, indicating the Omicron mutations favor the escape of current vaccines. 
In contrast, Delta's distribution focuses on a smaller domain as shown in 
Figure  \ref{fig:combine2}{\bf d}. 
 
Among 15 RBD mutations, K417N, also part of the Beta variant that originated in South Africa, causes the most significant disruption of known antibodies.   
Notably, E484A is another mutation that leads to overwhelmingly disruptive effects to many known antibodies. It is worthy of mentioning that most of E484A's disruptive effects are complementary to those of K417N, which makes Omicron more effective in vaccine breakthroughs. The third disruptive mutation is Y505H. It is also able to weaken many known antibody and RBD complexes. 

Mutation G339D creates a mild impact on various antibody-RBD complexes. One of the reasons is that it locates pretty far away from the binding interfaces of most known antibodies. Its change from a non-charged amino acid to a negatively charged amino acid induces mostly favorable bindings among many antibody-RBD complexes. 
S371L, S373P, and  S375F are other mutations that have mild impacts due to their locations. 

For a comparison, ACE2 is also included in Figure \ref{fig:combine2}. The impact of Omicron on ACE2 is significantly weak, indicating the SARS-CoV-2 has already optimized its binding with ACE2, and there is a relatively limited potential for the virus to improve its infectivity. However, due to the increase in the vaccination rate, variants can become more destructive to vaccines in years to come \cite{wang2021mechanisms}. 

It becomes very subtle to judge whether a mutation would disrupt an antibody and RBD complex as  Omicron involves multiple RBD mutations, which may generate multiple cancellations for each antibody and RBD complex over different mutations. Previously, we have used -0.3 kcal/mol as a threshold to judge whether a mutation disrupts an antibody and RBD complex, which would give us a total of 119   disrupted antibody and RBD complexes as shown in Figure \ref{fig:combine2}{\bf e}, suggesting a rate of 0.9 (i.e., 119/132) for potential vaccine breakthrough. 
As a comparison, Delta has 46 counts and a rate of 0.35 (46/132) in a similar estimation as shown in Figure \ref{fig:combine2}{\bf f}. One would have 108 and 30 disrupted antibody and RBD complexes respectively for Omicron and Delta if the threshold is increased to -0.6kcal/mol. Only two antibodies and RBD complexes have smaller than -3 kcal/mol mutation-induced BFE changes. If the cancellations between positive and negative changes are considered,  we would have 77  complexes that have accumulated BFE changes smaller than  -0.3 kcal/mol, suggesting an antibody disruption rate of 0.58 (i.e., 77/132) for Omicron,  compared to 0.28 (37/132) for Delta. As mentioned earlier, the actual rate of Omicron-induced vaccine breakthrough might never be properly determined for the world population. 
  
Figure \ref{fig:combine2}{\bf a} gives a separated plot of the impacts of Omicron on a few mAbs. Similarly, there are dramatic reductions in their efficacy. A more specific discussion is given in the next section.  

\subsection{Antibody resistance} 

The assessment of Omicron's mutational threats to FDA-approved mAbs and a few other mAbs in clinical development is of crucial importance. Our AI-based predictions of similar threats from other variants, namely Alpha, Beta, Gamma, Delta, Epsilon, and Kappa, have shown an excellent agreement with experimental data \cite{chen2021revealing}. In this section, we focus on a few mAbs, specifically,  mAbs from Eli Lilly (LY-CoV016 and LY-CoV555) and Regeneron (REGN10933, REGN10987, and REGN10933/10987), one mAb from Celltrion (CT-P59), two mAbs from the Rockefeller University (C135 and C144). Among them, mAbs from Eli Lilly  and Regeneron have had FDA approval. In addition, Celltrion's COVID-19 antibody treatment had the EU drug agency's recommendation in November 2021. Rockefeller University's mAbs are still in clinical trials. 
\begin{figure}[ht!]
	\centering
	\includegraphics[width = 0.95\textwidth]{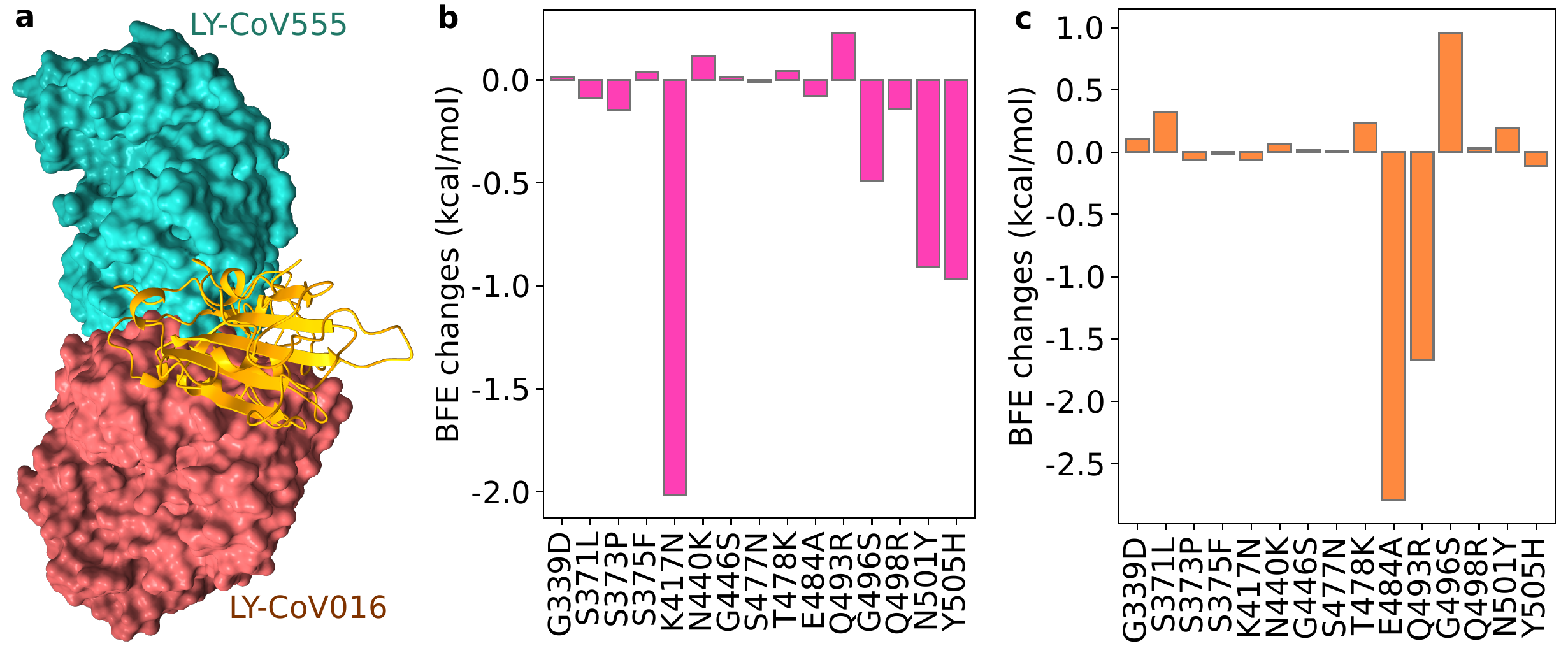}
	\caption{Illustration of the Omicron RBD and Eli Lilly antibody interaction and RBD mutation-induced BFE changes. 
{\bf a} The 3D structure of the ACE2 and Eli Lilly antibody  complex.  LY-CoV555 (PDB ID: 7KMG\cite{jones2021neutralizing}) and LY-CoV016 (PDB ID: 7C01\cite{shi2020human}) overlap on the S protein RBD. 
{\bf b} Omicron mutation-induced BFE changes for the complex of RBD and LY-CoV016.  
{\bf c} Omicron mutation-induced BFE changes for the complex of RBD and LY-CoV555.
}
		\label{fig:combine3}
\end{figure}

\paragraph{Eli Lilly mAbs}
Eli Lilly mAb LY-CoV555 (PDB ID: 7KMG\cite{jones2021neutralizing}) is also known as Bamlanivimab and is used in combination with LY-CoV016  (aka Etesevimab, PDB ID: 7C01\cite{shi2020human}). Antibody LY-CoV016 is isolated from patient  peripheral blood mononuclear cells convalescing from COVID-19. It was optimized based on the SARS-CoV-2 virus. The interaction of Eli Lilly mAbs  with the S protein RBD is depicted in Figure \ref{fig:combine3}{\bf a}. Clearly, LY-CoV555 has a competing relationship with LY-CoV016, which might complicate our predictions slightly. In this work, we carry out the analysis of  Eli Lilly mAbs separately. 

Omicron mutation-induced BFE changes for antibody LY-CoV016 and RBD complex is given in Figure~\ref{fig:combine3}{\bf b}. It appears that LY-CoV555 was  optimized with respect to the original S protein but is sensitive to mutations. This complex may be weakened by K417N and N501Y as predicted in our earlier work \cite{chen2021revealing}. New mutation Y505H may also reduce LY-CoV016's efficacy. Overall, the complex may be significantly weakened by Omicron, leading to the efficacy reduction of  Etesevimab. 

The predicted BFE changes of LY-CoV555 are shown in Figure~\ref{fig:combine3}{\bf c}. Mutation E484A  induces a negative BFE change of -2.79 kcal/mol for the LY-CoV555 and RBD complex.  The BFE change may translate into a dramatic efficacy reduction of 16 times for LY-CoV555, making it less competitive with ACE2 as most Omicron mutations strengthen the S protein and ACE2 binding. Similarly, Q493R may also reduce the efficacy by about 5 times. However, G496S may enhance the binding of the complex by 2.6 times. The impacts of other mutations are mild. Therefore, Omicron is expected to reduce LY-CoV555 efficacy significantly. A previous study indicated that LY-CoV555 is prone to the E484K mutation presented in Beta and Gamma variants, for which the Eli Lilly mAb cocktail was taken off the market for many months in 2021.

Although LY-CoV555 and LY-CoV016 might slightly complement, they are both prone to Omicron mutation-induced efficacy reduction. We predict that the Eli Lilly mAb cocktail would be retaken off the market had Omicron become a prevailing variant in the world. 
 
\begin{figure}[ht!]
\centering
\includegraphics[width = 0.95\textwidth]{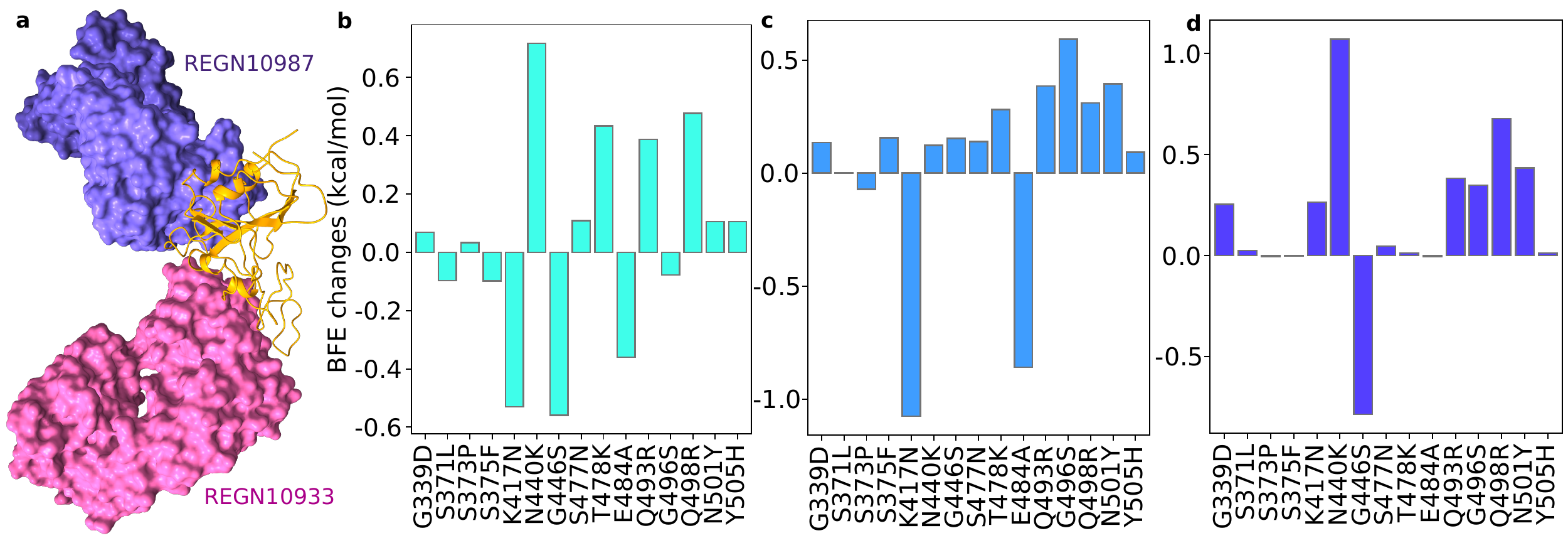}
\caption{Illustration of the Omicron RBD and Regeneron antibody interaction and RBD mutation-induced BFE changes. 
{\bf a} The 3D structure of the ACE2 and  Regeneron  antibody  complex.  REGN10987 and REGN10933 do not overlap on the S protein RBD (PDB ID: 6XDG\cite{hansen2020studies}). 
{\bf b} Omicron mutation-induced BFE changes for the complex of RBD and REGN10933.  
{\bf c} Omicron mutation-induced BFE changes for the complex of RBD and REGN10987.  
{\bf d} Omicron mutation-induced BFE changes for the complex of RBD, REGN10933, and REGN10987.  
}
\label{fig:combine4}
\end{figure}

\paragraph{Regeneron mAbs}

Regeneron mAbs REGN10933 and REGN10987 (aka Casirivimab and Imdevimab, respectively)
are an FDA-approved antibody cocktail (PDB ID: 6XDG\cite{hansen2020studies}) against COVID-19. Their 3D structure in complex with the S protein RBD is depicted in Figure \ref{fig:combine4}{\bf a}. 
Unlike the Eli Lilly mAb cocktail, the Regeneron mAbs do not overlap each other 
and bind to different parts of the RBD. Our 3D alignment shows that the antibody REGN10987 does not directly compete with ACE2 \cite{chen2021revealing}, which suggests that REGN10933 plays a more critical role in its ability to neutralize the virus directly.

Figure \ref{fig:combine4}{\bf b} plots our AI predicted BFE changes of the REGN10987-RBD complex. There are mixed responses to various Omicron mutations. Although G446K  and K417N induce a negative BFE change, many other mutations enhance the binding of the complex. We expect the overall binding strengthening effect. 

Omicron-induced BFE changes of the REGN10933-RBD complex is given in Figure \ref{fig:combine4}{\bf c}. Apparently, K417N and E484A  induce BFE changes of -1.08 and -0.86 kcal/mol, respectively. However, most other Omicron mutations strengthen the  binding of the complex. Particularly, G496S induces a BFE change of 0.55 kcal/mol. These two opposite effects may neutralize each other. Therefore, we predict very mild changes in  Casirivimab's efficacy against COVID-19.

It is interesting to study how the two Regeneron mAbs are affected by Omicron when they are combined. Figure \ref{fig:combine4}{\bf d} shows the BFE changes of the complex induced by various Omicron mutations. We note that amplitudes of both positive and negative BEF changes have significantly reduced, indicating a robust response of the cocktail against Omicron RBD mutations. We predict that Omicron will have a very mild negative impact on the  Regeneron cocktail if it does not enhance the its efficacy. 

\paragraph{The mAbs from Celltrion and Rockefeller University}

Celltrion's antibody  CT-P59 (aka Regdanvimab, PDB ID: 7CM4) is used a  cocktail with CT-P63, for which we do not have its 3D structure. Antibody CT-P59 binds the RBD in a completing region with ACE2 (see the Supporting Information) and thus, might play a more important role than CT-P63 in combating the virus.  Figure S1 {\bf b} shows that mutations E484A, Q493R, and Q498R respectively  lead to  BFE changes of -1.49, -2.82, and -1.0 kcal/mol for the CT-P59-RBD complex. These disruptive effects are offset by a positive BFE change of 1.71 kcal/mol due to mutation N501Y, which was reported in our earlier work \cite{chen2021revealing}.  The impacts of other mutations are relatively mild. Previously, we have shown that  CT-P59 is prone to L452R in Delta and Q439R and S494P \cite{chen2021revealing}. Due to the lack of the CT-P63 structure, we cannot provide an inclusive estimation for Celltrion's cocktail but would recommend caution toward the use of Celltrion's  Regdanvimab in the wake of Omicron infections.

Finally, we analyze Rockefeller University antibodies C135 (PDB ID: 7K8Z) and C144 (PDB ID: 7K90), whose binding complexes with the RBD are given in the Supporting Information. Antibody C135 has a relatively small region of interface with RBD. 
Our earlier study indicates that  C135  is prone to R346K and R346S mutations, while the efficacy of C144 can be significantly  reduced E484K in the Delta variant. Figure S1 {\bf e} shows that most of Omicron's mutations, except for S317L,  have a mild impact on antibody C135, with their amplitudes being small than 0.5 kcal/mol. Mutation S317L induces a BFE change of -0.63 kcal/mol, indicating a relatively weak negative impact on C135's efficacy.  

In contrast, antibody C144 has more dramatic responses to Omicron mutations (see Figure S1 {\bf f}). Mutation E484A may cause a BFE change of -1.27kcal/mol. All other mutations may not affect C144 very much. Therefore, we predict that the efficacy of C144 may  be also undermined by Omicron RBD mutations.

\section{Data, methods and validity}
To deliver an accurate and reliable machine learning model, dataset collection is of paramount importance among other steps. Both the BFE changes and next-generation sequencing enrichment ratios indicate the mutation-induced effects on protein-protein interactions (PPIs) binding affinities. Our methods integrate these two types of datasets to improve the prediction accuracy \cite{chen2021prediction,chen2021revealing}. Considering the urgency of COVID-19, the scattered SARS-CoV-2 data concerning BFE changes are reported inconsistently, while the sequencing enrichment ratios data is relatively easy to obtain but consistently has particular protein-protein interaction problems. The method is set up based on the BFE change dataset, SKEMPI 2.0 \cite{jankauskaite2019skempi}, together with SARS-CoV-2 related datasets. These datasets are obtained from the mutational scanning on ACE2 binding to the S protein RBD \cite{chan2020engineering}, the mutational scanning on RBD binding to ACE2 \cite{starr2020deep,linsky2020novo}, and the mutational scanning on RBD binding to CTC-445.2 and on CTC-445.2 binding to RBD\cite{linsky2020novo}. We have also collected a library of 132 3D structures of  antibody-RBD complexes \cite{chen2021revealing}. 

Our deep learning model for predicting BFE changes induced by mutations is constructed in two main steps. Firstly, once 3D structures of PPI complexes are obtained, mathematical features and biochemical/biophysical features are extracted. Biochemical/biophysical features provide the chemical and physical information, such as surface areas, partial charges, Coulomb interactions, van der Waals interaction, electrostatics, etc. Mathematical features, including the element-specific and site-specific persistent homology (algebraic topology), are implemented to simplify the structural complexity of PPI complexes \cite{chen2020mutations,wang2020topology}. Second, a deep learning algorithm, artificial neural networks (ANNs), is constructed to tackle the massive features and mutational scanning data for predictions \cite{chen2021revealing}, which is available at \href{https://github.com/WeilabMSU/TopNetmAb}{TopNetmAb}. Notice that our early model was constructed by integrating convolutional neural networks (CNNs) with gradient boosting trees (GBTs) and was trained with a large dataset of 8,338 PPI entries from the SKEMPI 2.0 dataset \cite{jankauskaite2019skempi}, which had already achieved a high accuracy \cite{wang2020topology}. 

In more recent work \cite{chen2021prediction,chen2021revealing,wang2021emerging}, with the help of the aforementioned deep mutational datasets associated with SARS-CoV-2, our predictions are highly consistent with experimental data. The predictions for the binding of CTC-445.2 and S protein RBD were compared with experimental data with a Pearson correlation of 0.7 \cite{chen2021revealing,linsky2020novo}. As a baseline, one may keep in mind that experimental deep mutational results for SARS-COV-2 PPIs from 2 different labs  only have a correlation of 0.67 \cite{starr2020deep,linsky2020novo}.
In the same work \cite{chen2021revealing}, the predictions of emerging mutations on clinical trial antibodies had a Pearson correlation of 0.8 with the natural log of experimental escape fractions \cite{starr2021prospective}. In addition, the predicted mutation-induced BFE changes on L452R and N501Y for the ACE2-RBD complex have a near perfect correlation with experimental luciferase data \cite{chen2021revealing,deng2021transmission}.

Our TopNetmAb model assumes that the RBD mutations are independent, which is very reasonable for Delta and other variants as they involve only one, two, or three  RBD mutations. As shown in Figure \ref{fig:combine1}{\bf a}, adjacent Omicron mutations S477N and T478K are dependent on each other. Similarly, S373P is just one residue away from S371L and S375F and is deemed to be depending on its neighbors. However, these three mutations are pretty far away from the ACE2 binding interface and play a less important role in our predictions. Mutation G496S and Q498R are also one amino acid apart, albeit their predicted BFE change amplitudes are very small. Overall, we expect a larger error in our prediction of Omicron infectivity compared to our earlier successful predictions \cite{chen2021prediction,wang2021vaccine, chen2021revealing}. However, we are still confident that the predicted trend of the Omicron infectivity change is correct. 
Figure \ref{fig:combine2}{\bf b1-b2} shows that the most severe antibody disruptions are not obtained from the inter-dependent mutations (i.e., S371L, S373P, S375F, S477N,  T478K, G496S, and Q498R), suggesting the predicted trend of antibody disruptions is still valid. The reliability and accuracy of our assumption for Omicron are to be validated by experimental data, which may become available in a few weeks.

\section{Conclusion}

The identification of Omicron as a variant of concern (VOC) by the World Health Organization (WHO) has triggered countries around the world to put in place of travel restrictions and precautionary measures. At this moment, the scientific community knows  little about Omicron's infectivity, vaccine breakthrough, and antibody resistance. Since the spike (S) protein, particularly, its receptor-binding domain (RBD), plays a vital role in viral infection, it has been a key target of vaccines and antibody drugs. Therefore, the study of Omicron's 15 RBD mutations can lead to valuable understanding of Omicron's infectivity, vaccine breakthrough, and antibody resistance.

Based on a well-tested and experimentally confirmed deep learning model trained with tens of thousands of experimental data, we investigate the impacts of  Omicron's RBD mutations to its infectivity. We show that Omicron is about ten times more infectious than the original virus or about twice as infectious as the Delta variant. Using the structures of 132 known antibody-RBD complexes, we reveal that Omicron's vaccine-escape capability is about twice as high as that of the Delta variant. We unveil that Omicron may significantly reduce the efficacy of Eli Lilly antibody  cocktail. Omicron RBD mutations may also compromise monoclonal
antibodies (mAbs) from  Celltrion and Rockefeller University.

\section*{Data and model availability}

The structural information of 132 antibody-RBD complexes with their corresponding PDB IDs and the results of BFE changes of PPI complexes induced by Omicron mutations can be found in Section {\color{teal} S2} of the Supporting Information. The analysis of observed SARS-CoV-2 RBD mutations is available at \href{https://weilab.math.msu.edu/MutationAnalyzer/}{Mutaton Analyzer}. The TopNetTree model is available at \href{https://github.com/WeilabMSU/TopNetmAb}{TopNetmAb}. 
The detailed methods can be found in the Supporting Information {\color{teal} S3} and {\color{teal} S4}. The validation of our predictions with experimental data can be located in Supporting Information {\color{teal} S5}.

\section*{Supporting information}
The supporting information is available for 
\begin{enumerate}
    \item[S1] Supplementary figures: BFE changes of CT-P59, C135, and C144.
    \item[S2] Supplementary data: The Supplementary\_Data.zip contains two files: the BFE changes of antibodies disrupted by Omicron mutations and the list of antibodies with corresponding PDB IDs.
    \item[S3] Supplementary data pre-processing and feature generation methods
    \item[S4] Supplementary machine learning methods.
    \item[S5] Supplementary validation: validations of our machine learning predictions with experimental data.
\end{enumerate}

\section*{Acknowledgment}
This work was supported in part by NIH grant  GM126189, NSF grants DMS-2052983,  DMS-1761320, and IIS-1900473,  NASA grant 80NSSC21M0023,  Michigan Economic Development Corporation, MSU Foundation,  Bristol-Myers Squibb 65109, and Pfizer.


\begin{thebibliography}{10}
	
	\bibitem{li2005bats}
	Wendong Li, Zhengli Shi, Meng Yu, Wuze Ren, Craig Smith, Jonathan~H Epstein,
	Hanzhong Wang, Gary Crameri, Zhihong Hu, Huajun Zhang, et~al.
	\newblock Bats are natural reservoirs of {SARS}-like coronaviruses.
	\newblock {\em Science}, 310(5748):676--679, 2005.
	
	\bibitem{qu2005identification}
	Xiu-Xia Qu, Pei Hao, Xi-Jun Song, Si-Ming Jiang, Yan-Xia Liu, Pei-Gang Wang,
	Xi~Rao, Huai-Dong Song, Sheng-Yue Wang, Yu~Zuo, et~al.
	\newblock Identification of two critical amino acid residues of the severe
	acute respiratory syndrome coronavirus spike protein for its variation in
	zoonotic tropism transition via a double substitution strategy.
	\newblock {\em Journal of Biological Chemistry}, 280(33):29588--29595, 2005.
	
	\bibitem{song2005cross}
	Huai-Dong Song, Chang-Chun Tu, Guo-Wei Zhang, Sheng-Yue Wang, Kui Zheng,
	Lian-Cheng Lei, Qiu-Xia Chen, Yu-Wei Gao, Hui-Qiong Zhou, Hua Xiang, et~al.
	\newblock Cross-host evolution of severe acute respiratory syndrome coronavirus
	in palm civet and human.
	\newblock {\em Proceedings of the National Academy of Sciences},
	102(7):2430--2435, 2005.
	
	\bibitem{hoffmann2020sars}
	Markus Hoffmann, Hannah Kleine-Weber, Simon Schroeder, Nadine Kr{\"u}ger, Tanja
	Herrler, Sandra Erichsen, Tobias~S Schiergens, Georg Herrler, Nai-Huei Wu,
	Andreas Nitsche, et~al.
	\newblock {SARS-CoV-2} cell entry depends on {ACE2} and {TMPRSS2} and is
	blocked by a clinically proven protease inhibitor.
	\newblock {\em Cell}, 181(2):271--280, 2020.
	
	\bibitem{walls2020structure}
	Alexandra~C Walls, Young-Jun Park, M~Alejandra Tortorici, Abigail Wall,
	Andrew~T McGuire, and David Veesler.
	\newblock Structure, function, and antigenicity of the {SARS-CoV-2} spike
	glycoprotein.
	\newblock {\em Cell}, 2020.
	
	\bibitem{wang2020human}
	Chunyan Wang, Wentao Li, Dubravka Drabek, Nisreen~MA Okba, Rien van Haperen,
	Albert~DME Osterhaus, Frank~JM van Kuppeveld, Bart~L Haagmans, Frank
	Grosveld, and Berend-Jan Bosch.
	\newblock A human monoclonal antibody blocking {SARS-CoV-2} infection.
	\newblock {\em Nature communications}, 11(1):1--6, 2020.
	
	\bibitem{yu2020receptor}
	Fei Yu, Rong Xiang, Xiaoqian Deng, Lili Wang, Zhengsen Yu, Shijun Tian, Ruiying
	Liang, Yanbai Li, Tianlei Ying, and Shibo Jiang.
	\newblock Receptor-binding domain-specific human neutralizing monoclonal
	antibodies against {SARS-CoV} and {SARS-CoV-2}.
	\newblock {\em Signal Transduction and Targeted Therapy}, 5(1):1--12, 2020.
	
	\bibitem{li2021impact}
	Cheng Li, Xiaolong Tian, Xiaodong Jia, Jinkai Wan, Lu~Lu, Shibo Jiang, Fei Lan,
	Yinying Lu, Yanling Wu, and Tianlei Ying.
	\newblock The impact of receptor-binding domain natural mutations on antibody
	recognition of {SARS-CoV-2}.
	\newblock {\em Signal Transduction and Targeted Therapy}, 6(1):1--3, 2021.
	
	\bibitem{chen2020mutations}
	Jiahui Chen, Rui Wang, Menglun Wang, and Guo-Wei Wei.
	\newblock Mutations strengthened {SARS-CoV-2} infectivity.
	\newblock {\em Journal of molecular biology}, 432(19):5212--5226, 2020.
	
	\bibitem{chen2021prediction}
	Jiahui Chen, Kaifu Gao, Rui Wang, and Guo-Wei Wei.
	\newblock Prediction and mitigation of mutation threats to {COVID-19} vaccines
	and antibody therapies.
	\newblock {\em Chemical Science}, 12(20):6929--6948, 2021.
	
	\bibitem{chen2021revealing}
	Jiahui Chen, Kaifu Gao, Rui Wang, and Guo-Wei Wei.
	\newblock Revealing the threat of emerging {SARS-CoV-2} mutations to antibody
	therapies.
	\newblock {\em Journal of Molecular Biology}, 433(7744), 2021.
	
	\bibitem{alenquer2021signatures}
	Marta Alenquer, Filipe Ferreira, Diana Lousa, Mariana Val{\'e}rio, M{\'o}nica
	Medina-Lopes, Marie-Louise Bergman, Juliana Gon{\c{c}}alves, Jocelyne
	Demengeot, Ricardo~B Leite, Jingtao Lilue, et~al.
	\newblock Signatures in sars-cov-2 spike protein conferring escape to
	neutralizing antibodies.
	\newblock {\em PLoS pathogens}, 17(8):e1009772, 2021.
	
	\bibitem{dupont2021neutralizing}
	Liane Dupont, Luke~B Snell, Carl Graham, Jeffrey Seow, Blair Merrick, Thomas
	Lechmere, Thomas~JA Maguire, Sadie~R Hallett, Suzanne Pickering, Themoula
	Charalampous, et~al.
	\newblock Neutralizing antibody activity in convalescent sera from infection in
	humans with sars-cov-2 and variants of concern.
	\newblock {\em Nature microbiology}, pages 1--10, 2021.
	
	\bibitem{wang2021vaccine}
	Rui Wang, Jiahui Chen, Kaifu Gao, and Guo-Wei Wei.
	\newblock Vaccine-escape and fast-growing mutations in the {United Kingdom, the
		United States, Singapore, Spain, India}, and other {COVID}-19-devastated
	countries.
	\newblock {\em Genomics}, 113(4):2158--2170, 2021.
	
	\bibitem{linsky2020novo}
	Thomas~W Linsky, Renan Vergara, Nuria Codina, Jorgen~W Nelson, Matthew~J
	Walker, Wen Su, Christopher~O Barnes, Tien-Ying Hsiang, Katharina
	Esser-Nobis, Kevin Yu, et~al.
	\newblock De novo design of potent and resilient {hACE2} decoys to neutralize
	{SARS-CoV-2}.
	\newblock {\em Science}, 370(6521):1208--1214, 2020.
	
	\bibitem{zhang2021furin}
	Liping Zhang, Matthew Mann, Zulfeqhar Syed, Hayley~M Reynolds, E~Tian, Nadine~L
	Samara, Darryl~C Zeldin, Lawrence~A Tabak, and Kelly~G Ten~Hagen.
	\newblock Furin cleavage of the sars-cov-2 spike is modulated by
	o-glycosylation.
	\newblock {\em Proceedings of the National Academy of Science of the United
		States of America}, 118(47):e2109905118, 2021.
	
	\bibitem{wang2021emerging}
	Rui Wang, Jiahui Chen, Yuta Hozumi, Changchuan Yin, and Guo-Wei Wei.
	\newblock Emerging vaccine-breakthrough {SARS-CoV-2} variants.
	\newblock {\em arXiv preprint arXiv:2103.08023}, 2021.
	
	\bibitem{chen2021review2}
	Jiahui Chen, Rui Wang, and Guo-Wei Wei.
	\newblock Review of the mechanisms of sars-cov-2 evolution and transmission.
	\newblock {\em ArXiv}, 2021.
	
	\bibitem{callaway2020race}
	Ewen Callaway.
	\newblock The race for coronavirus vaccines: a graphical guide.
	\newblock {\em Nature}, 580(7805):576, 2020.
	
	\bibitem{dai2021viral}
	Lianpan Dai and George~F Gao.
	\newblock Viral targets for vaccines against covid-19.
	\newblock {\em Nature Reviews Immunology}, 21(2):73--82, 2021.
	
	\bibitem{wang2021mechanisms}
	Rui Wang, Jiahui Chen, and Guo-Wei Wei.
	\newblock Mechanisms of sars-cov-2 evolution revealing vaccine-resistant
	mutations in europe and america.
	\newblock {\em The Journal of Physical Chemistry Letters}, 2021.
	
	\bibitem{jones2021neutralizing}
	Bryan~E Jones, Patricia~L Brown-Augsburger, Kizzmekia~S Corbett, Kathryn
	Westendorf, Julian Davies, Thomas~P Cujec, Christopher~M Wiethoff, Jamie~L
	Blackbourne, Beverly~A Heinz, Denisa Foster, et~al.
	\newblock The neutralizing antibody, ly-cov555, protects against sars-cov-2
	infection in nonhuman primates.
	\newblock {\em Science translational medicine}, 13(593), 2021.
	
	\bibitem{shi2020human}
	Rui Shi, Chao Shan, Xiaomin Duan, Zhihai Chen, Peipei Liu, Jinwen Song, Tao
	Song, Xiaoshan Bi, Chao Han, Lianao Wu, et~al.
	\newblock A human neutralizing antibody targets the receptor binding site of
	{SARS-CoV-2}.
	\newblock {\em Nature}, pages 1--8, 2020.
	
	\bibitem{hansen2020studies}
	Johanna Hansen, Alina Baum, Kristen~E Pascal, Vincenzo Russo, Stephanie
	Giordano, Elzbieta Wloga, Benjamin~O Fulton, Ying Yan, Katrina Koon, Krunal
	Patel, et~al.
	\newblock Studies in humanized mice and convalescent humans yield a
	{SARS-CoV-2} antibody cocktail.
	\newblock {\em Science}, 369(6506):1010--1014, 2020.
	
	\bibitem{jankauskaite2019skempi}
	Justina Jankauskait{\.e}, Brian Jim{\'e}nez-Garc{\'\i}a, Justas Dapk{\=u}nas,
	Juan Fern{\'a}ndez-Recio, and Iain~H Moal.
	\newblock {SKEMPI} 2.0: an updated benchmark of changes in protein--protein
	binding energy, kinetics and thermodynamics upon mutation.
	\newblock {\em Bioinformatics}, 35(3):462--469, 2019.
	
	\bibitem{chan2020engineering}
	Kui~K Chan, Danielle Dorosky, Preeti Sharma, Shawn~A Abbasi, John~M Dye,
	David~M Kranz, Andrew~S Herbert, and Erik Procko.
	\newblock Engineering human {ACE2} to optimize binding to the spike protein of
	{SARS} coronavirus 2.
	\newblock {\em Science}, 369(6508):1261--1265, 2020.
	
	\bibitem{starr2020deep}
	Tyler~N Starr, Allison~J Greaney, Sarah~K Hilton, Daniel Ellis, Katharine~HD
	Crawford, Adam~S Dingens, Mary~Jane Navarro, John~E Bowen, M~Alejandra
	Tortorici, Alexandra~C Walls, et~al.
	\newblock Deep mutational scanning of {SARS-CoV-2} receptor binding domain
	reveals constraints on folding and {ACE2} binding.
	\newblock {\em Cell}, 182(5):1295--1310, 2020.
	
	\bibitem{wang2020topology}
	Menglun Wang, Zixuan Cang, and Guo-Wei Wei.
	\newblock A topology-based network tree for the prediction of protein--protein
	binding affinity changes following mutation.
	\newblock {\em Nature Machine Intelligence}, 2(2):116--123, 2020.
	
	\bibitem{starr2021prospective}
	Tyler~N Starr, Allison~J Greaney, Amin Addetia, William~W Hannon, Manish~C
	Choudhary, Adam~S Dingens, Jonathan~Z Li, and Jesse~D Bloom.
	\newblock Prospective mapping of viral mutations that escape antibodies used to
	treat {COVID-19}.
	\newblock {\em Science}, 371(6531):850--854, 2021.
	
	\bibitem{deng2021transmission}
	Xianding Deng, Miguel~A Garcia-Knight, Mir~M Khalid, Venice Servellita, Candace
	Wang, Mary~Kate Morris, Alicia Sotomayor-Gonz{\'a}lez, Dustin~R Glasner,
	Kevin~R Reyes, Amelia~S Gliwa, et~al.
	\newblock Transmission, infectivity, and antibody neutralization of an emerging
	{SARS-CoV-2} variant in {C}alifornia carrying a {L452R} spike protein
	mutation.
	\newblock {\em MedRxiv}, 2021.
	
\end{thebibliography}
\end{document}